# How to measure work functions from aqueous solutions

Michele Pugini[1], Bruno Credidio[1], Irina Walter[1], Sebastian Malerz[1], Florian Trinter[1,2], Dominik Stemer[1], Uwe Hergenhahn[1], Gerard Meijer[1], Iain Wilkinson[3], Bernd Winter[1]*, and Stephan Thürmer[4]*

[1] *Fritz-Haber-Institut der Max-Planck-Gesellschaft, Faradayweg 4-6, 14195 Berlin, Germany*
[2] *Institut für Kernphysik, Goethe-Universität, Max-von-Laue-Straße 1, 60438 Frankfurt am Main, Germany*
[3] *Institute for Electronic Structure Dynamics, Helmholtz-Zentrum Berlin für Materialien und Energie, Hahn-Meitner-Platz 1, 14109 Berlin, Germany*
[4] *Department of Chemistry, Graduate School of Science, Kyoto University, Kitashirakawa-Oiwakecho, Sakyo-Ku, 606-8502 Kyoto, Japan*

ORCID
MP: 0000-0003-2406-831X
BC: 0000-0003-0348-0778
I Walter: 0000-0001-8060-4143
SM: 0000-0001-9570-3494
FT: 0000-0002-0891-9180
DS: 0000-0002-5528-1773
UH: 0000-0003-3396-4511
GM: 0000-0001-9669-8340
I Wilkinson: 0000-0001-9561-5056
BW: 0000-0002-5597-8888
ST: 0000-0002-8146-4573

* Corresponding authors: winter@fhi-berlin.mpg.de; thuermer@kuchem.kyoto-u.ac.jp

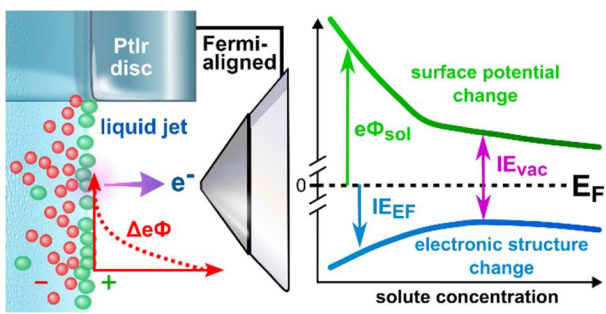

Fermi-referencing and work-function determination from aqueous solutions is enabled by the control of extrinsic potentials, which are unique to streaming liquids. Concentration-dependent changes in both quantities are described for the first time.

# Abstract

The recent application of concepts from condensed-matter physics to photoelectron spectroscopy (PES) of volatile, liquid-phase systems has enabled the measurement of electronic energetics of liquids on an absolute scale. Particularly, vertical ionization energies, VIEs, of liquid water and aqueous solutions, both in the bulk and at associated interfaces, can now be accurately, precisely, and routinely determined. These IEs are referenced to the local vacuum level, which is the appropriate quantity for condensed matter with associated surfaces, including liquids. In this work, we connect this newly accessible energy level to another important surface property, namely, the solution work function, $e\Phi_{liq}$. We lay out the prerequisites for and unique challenges of determining $e\Phi$ of aqueous solutions and liquids in general. We demonstrate – for a model aqueous solution with a tetra-*n*-butylammonium iodide (TBAI) surfactant solute – that concentration-dependent work functions, associated with the surface dipoles generated by the segregated interfacial layer of TBA$^+$ and I$^-$ ions, can be accurately measured under controlled conditions. We detail the nature of surface potentials, uniquely tied to the nature of the flowing-liquid sample, which must be eliminated or quantified to enable such measurements. This allows us to refer aqueous-phase spectra to the Fermi level and to quantitatively assign surfactant-concentration-dependent spectral shifts to competing work function and electronic-structure effects, where the latter are typically associated with solute–solvent interactions in the bulk of the solution which determine, *e.g.*, chemical reactivity. The present work describes the extension of liquid-jet PES to quantitatively access concentration-dependent surface descriptors that have so far been restricted to solid-phase measurements. Correspondingly, these studies mark the beginning of a new era in the characterization of the interfacial electronic structure of aqueous solutions and liquids more generally.

# Introduction

The addition of solutes to liquids generally leads to changes of solute and solvent polarization, both at the solution interface and in the bulk. Interfacial polarization and charge imbalance generates liquid-surface potentials,[1-4] which can affect interfacial chemical reactivity,[5-8] and notably offset the energetics of photoelectrons detected *via* such interfaces.[9-12] Furthermore, interfacial solute–solvent interactions can significantly affect the nature of surface electronic states,[13] further perturbing interfacial chemical phenomena. In the bulk of a solution, solute–solvent interactions and associated polarization may lead to energetic shifts of liquid-phase electronic states,[14, 15] which affect the chemical processes that occur within liquids. The experimental capability to separate and quantify liquid-phase potentials and electronic-structure changes correspondingly has the potential to provide deep, additional insights into the driving forces behind liquid-phase chemistry. With its variable sensitivity to interfacial and bulk-solution electronic structure[16] and related interfacial potentials,[17] photoelectron spectroscopy (PES) is well suited for the extraction of such chemically relevant electronic-structure information. We report a further conceptual step towards such measurements here. Recent advances in liquid-jet photoelectron spectroscopy (LJ-PES) have enabled the quantification and interpretation of energetic shifts of solvent (water) photoemission peaks upon the addition of solutes.[9-11, 15] These spectral effects have been attributed to changes of the solvent electron binding energies, eBEs, or ionization energies, IEs,[N1] which are central determiners of water-based chemistry. It has been demonstrated that the measurement of valence PE spectra and corresponding low-energy-cutoff positions, $E_{cut}$, can be used to accurately determine IEs of both the solvent and solute independently of an external reference, provided the ionizing photon energy, hv, is precisely known.[9, 15] In the present work, our focus is on the vertical IE (VIE), which is determined from the energetic positions of maximum intensity of discrete PE bands. Associated values correspond to the minimum energy required to release photoelectrons into vacuum without concurrent geometric structural rearrangement.[N2] IEs are then referenced to the solution's vacuum level, $E_{vac}$, which we denote as $IE_{vac}$, or $VIE_{vac}$ in the case of the specific vertical value of this quantity. The determination of accurate $IE_{vac}$ values was previously elusive, as energies could only be determined relative to some reference photoelectron peak. In the case of aqueous solutions, the references have usually been neat liquid water's lowest IE peak[18, 19] – commonly attributed to electron liberation from its $1b_1$ highest occupied molecular orbital (HOMO) – or its O 1s core-level[20] ionization feature. Such practices imply that any solute-induced changes of the water-solvent electronic structure, such as shifts of the $1b_1$ or O 1s reference bands themselves, were inaccessible and deemed negligible. Our recent results have shown that previously determined external-reference-reliant IEs were prone to errors, with solute IE values found to be off in some, so far limited, cases by several hundred meV.[9, 15] IEs of both solvent and solute typically depend on the specific solution system and component concentrations. In our previous study,[15] our energy-referencing procedure was applied to aqueous solutions of surface-active tetra-*n*-butylammonium iodide (TBAI) and highly soluble sodium iodide (NaI) for a large range of concentrations, extending close to the respective saturation limits. For TBAI$_{(aq)}$ solutions, a large energy shift of up to 700 meV towards lower $VIE_{vac}$ values was observed for both the water-solvent and solute PE peaks. Considerably smaller energetic shifts of 270 meV in the opposite direction were measured for the NaI$_{(aq)}$ solutions. Although those experiments allowed accurate $IE_{vac}$ values to be determined, the origin of the observed energy shifts, *i.e.*, whether they were caused by surface effects or by modifications to the interfacial/bulk-solution electronic structure, could not be unambiguously determined.[N3] To make such a distinction, the introduction of a different, absolute energy reference, namely the Fermi level, $E_F$, is required. $E_F$, formally equivalent to the electrochemical potential, $\bar{\mu}$, assumes the same energetic position throughout all matter in electrical contact and at thermodynamic equilibrium.[21] Thus, $E_F$ is usually a preferred energy reference in condensed-phase PE spectroscopy. Together with $E_{vac}$, this energy reference allows the solution's work function, WF or $e\Phi$, to be measured and in turn can give access to explicit interfacial descriptors such as solute surface enrichment, molecular orientation, and possible surface-potential inhomogeneities. Importantly, the combined energy-referencing schemes allow $e\Phi$ changes to be distinguished from possible variations in bulk-solution electronic structure. In our previous work,[9] we have already accurately determined $e\Phi$ values of liquid water and estimated them for TBAI$_{(aq)}$ solutions at a specific concentration of 25 mM. Here, we focus on the extension of this methodology to TBAI$_{(aq)}$ solutions of arbitrary concentration. We also provide an in-depth discussion of the intricacies of energy referencing in LJ-PES, with a perspective to generalize our methodology for application to arbitrary liquids.

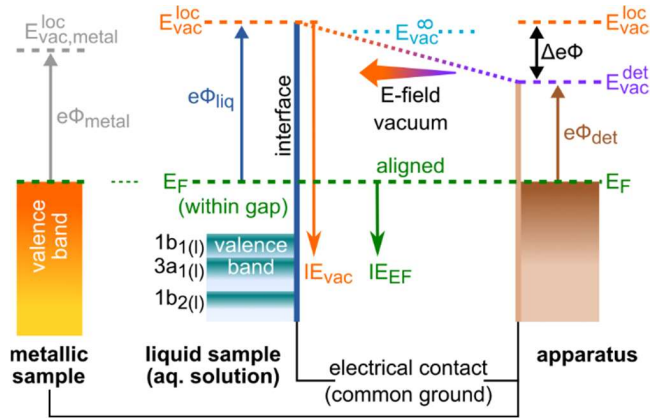

**Figure 1:** Schematic of the relevant energy levels in the $E_F$-referenced, *i.e.*, grounded and equilibrated, LJ-PES experiment. The left side depicts the sample, either a solid metallic reference or an aqueous-solution sample, each with a specific sample work function, $e\Phi_{liq}$. The right side represents the apparatus, with, most prominently, the detector entrance orifice positioned in the vicinity of the sample, with its own work function, $e\Phi_{det}$. In order to equalize the Fermi level, $E_F$, throughout the experimental components, the sample is brought in electrical contact with the apparatus. In the case of a metallic sample, the occupied states extend up to $E_F$. For an aqueous solution, $E_F$ usually resides in the band gap, devoid of available electronic states; the position of $E_F$ thus cannot be observed in a solution-phase PE spectrum. The work function connects $E_F$ to the local vacuum level, $E_{vac}^{loc}$, which represents the vacuum level just outside of the material's surface, as described in the main body of the text. For completeness, the vacuum level at infinity, $E_{vac}^{\infty}$, is the vacuum level far away from any matter. This is the reference level for gas-phase ionization energies, which is not of relevance for Fermi referencing. Because of the difference in work functions between the sample and the detector assembly, and in turn their local vacuum levels, a contact potential difference, $\Delta e\Phi = e\Phi_{liq} - e\Phi_{det}$, exists between the sample and apparatus, which results in an electric field that affects all photoelectrons; here, the arrow to the left indicates an accelerating field for the electrons. IE$_{vac}$ and IE$_{EF}$ are ionization-energy scales referenced to $E_{vac}^{loc}$ and $E_F$, respectively.

In order to experimentally disentangle bulk electronic structure[N4] from surface-related effects, the work function, $e\Phi$, needs to be made accessible to measurement, which inevitably requires the introduction of $E_F$. $e\Phi$ is defined as the minimum energy required to promote an electron with an IE equivalent to $E_F$ (of the bulk aqueous solution) into the vacuum. Hence, $e\Phi$ connects $E_F$ to the *local* vacuum level (detailed below), *i.e.*, $e\Phi = E_{vac}^{loc} - E_F$; see the energy-level diagram in Fig. 1. Note that $E_F$ is the intrinsic energy reference for the bulk electronic structure, and that electronic-structure changes reveal themselves as energy shifts of solute and/or solvent ionization features with respect to $E_F$. Changes of $e\Phi$, and associated energy shifts, on the other hand, result from modifications of the liquid's surface dipole layer, *i.e.,* the potential barrier for escaping photoelectrons. Both changes in $e\Phi$ and in the solution electronic structure lead to a variation of IE$_{vac}$ (unless the two effects cancel fortuitously), as observed in our previous studies.[9, 15] Without measuring the energy of the Fermi level, these two contributions cannot be meaningfully distinguished, let alone quantified. In this work, we present a protocol to disentangle the contributions of electronic-structure and $e\Phi$ solution properties to the changes in $IE_{vac}$, and demonstrate its application to the previously studied $TBAI_{(aq)}$ surfactant model system[9, 12, 22-25] over a range of solute concentrations. While we lay out a general protocol to determine bulk-electronic-structure effects, we note that in the present case of a surfactant solution, the bulk and interfacial electronic structure is necessarily different.[N4]

While there have been previous attempts to determine solution work functions, performed with specific solute concentrations,[10, 11, 26] the $E_F$ measurements were indirect (*e.g.*, the Fermi-referenced energy scale was determined in a roundabout way *via* the analyzer work function, $e\Phi_{det}$) and suffered from a number of issues, which we have previously discussed.[9] Another recent study determined $VIE_{vac}$ changes of various aqueous solutions from $E_{cut}$ and valence PES measurements, assigning the measured IE effects to work-function changes.[12] However, solution Fermi-level energies were not measured in that study and the observed $VIE_{vac}$ changes were instead inferred to relate to solute interfacial dipole-potential changes.

## Conceptual Design of our Method

The Fermi level, $E_F$, is inaccessible in semiconductors, which formally include aqueous solutions and liquids, since the respective band gap is devoid of available states for electrons. Instead, $E_F$ is a 'virtual' level, governed by the balance of available charge carriers in the valence and conduction bands, and must be determined indirectly.[27] The energetic position of $E_F$ which is constant for a given, electrically grounded apparatus, is first determined using a metallic reference sample (typically gold) and this reference energy is used to calibrate the spectra from the sample of interest, *i.e.*, the solution in our case.[27] The implicit assumption made here is that $E_F$ is aligned across the apparatus – *i.e.,* $E_F$ is equivalent for the reference metal, the sample of interest, and the electron detector – when all associated parts are properly grounded. For this referencing to work, it must be ensured that undesired extrinsic potentials – *i.e.*, potentials imposed by factors unrelated to Fermi alignment, which may alter the kinetic energy of the photoelectrons emitted from the sample – are eliminated or numerically compensated between the to-be-referenced sample and the electron detector. We stress that Fermi-referencing PE spectra from semiconductor materials always implicitly relies on the validity of the aforementioned assumption and condition. In practice, this may not be trivial to accomplish, even for solid samples such as semiconductors; photoemission can lead to sample charge-up if the sample conductivity is low and additional surface charge may be introduced *via* surface contamination, disturbing interfacial electrical equilibria. If such deleterious potentials are present, the measured spectrum will be shifted with respect to the previously determined $E_F$ position and thus the sample spectrum cannot be reliably energy-referenced.

Photoelectrons experience the sum of all acting potentials, which we will refer to as the total potential, $V_{tot}$, in the following, as illustrated in Fig. 2A. We discern two components of $V_{tot}$: The contact potential difference between sample and apparatus (also known as the Volta potential), $\Delta e\Phi$, and other (undesirable) *extrinsic* potentials, such as the streaming potential created by electrokinetic charging of the flowing liquid,[19, 28] or charging upon ionization. Here, $\Delta e\Phi$ is an *intrinsic* property of the sample and apparatus. In thermodynamic equilibrium, $\Delta e\Phi$ is caused by the difference in work functions of the apparatus and the sample, $\Delta e\Phi = e\Phi_{liq} - e\Phi_{det}$, with $e\Phi_{det}$ being the work function of the detection system, which is usually constant and independent of the sample.[21, 29] In our case, $\Delta e\Phi$ is associated with the potential between the LJ surface and the detection system, as illustrated in Fig. 1 by the sloped potential. $\Delta e\Phi$ may change if the sample work function is altered, *e.g.*, upon the introduction of a solute or a change of its concentration. We will explicitly comment on situations where different potential contributions to $V_{tot}$ can be singled out and thus quantitatively determined.

A unique challenge for liquid-phase PE spectroscopy is that the liquid sample must be introduced into vacuum as a fast-flowing microjet, which is essential to prevent instantaneous freezing.[N5] The liquid flow disrupts the electric double layer at the contact surface between the liquid and the inner capillary walls of the injection orifice (Helmholtz layer[30, 31]), which in turn leads to electrokinetic charging of the jet and the build-up of the streaming potential, $\Phi_{str}$.[19, 28] The parameters that determine $\Phi_{str}$ include the type and amount of solute, flow rate *versus* injection-orifice diameter (*i.e.*, flow speed), and temperature.[28, 32-34] It is possible to 'tune' electrokinetic charging by adding specific (small) amounts of salt, which alters the composition of the Helmholtz layer. In some cases, addition of salt also accounts for unbalanced positive surface charge following ionization. $\Phi_{str}$ can assume values up to several volts in some cases,[19, 28] and its elimination or quantification is of paramount importance when attempting to measure Fermi-referenced PE spectra from solutions, as we will detail here.

The presence of undesirable potentials implies that a fixed position of $E_F$ in a *measured* spectrum cannot be readily assumed, even if the *intrinsic* $E_F$ (*i.e.*, the actual potential inside the sample) may be very well aligned with the apparatus. In fact, these additional potentials make it difficult to obtain Fermi-referenced PE spectra from solutions. Here, we show how conditions can be engineered to reliably associate a reference $E_F$ value with measured liquid-phase PE spectra, and in turn determine $e\Phi_{liq}$ values. Following the procedure outlined in our earlier work,[9] we recorded PE spectra from a liquid jet under what we refer to as 'streaming-potential-free' conditions, implying that all additional extrinsic potentials introduced by the flowing liquid stream are eliminated. It is also assured that ionization does not introduce charging. Notably, this does not include the elimination $\Delta e\Phi$, as its presence is crucial for a proper Fermi-level alignment, see Fig. 1. With some restrictions, this allows us to establish an $E_F$-referenced energy scale for the sample solutions. We correspondingly obtain the energetic positions of associated PE features with respect to the Fermi level, *i.e.*, $VIE_{EF}$, allowing solution work functions to be determined *via* $e\Phi = VIE_{vac} - VIE_{EF}$ (as illustrated in Fig. 3 later), notably as a function of solute concentration. Further pitfalls and considerations when attempting to measure Fermi-referenced PE spectra from LJs are summarized in the Supporting Information (SI).

We briefly comment on the two different vacuum levels shown in Fig. 1: The vacuum level relevant for gas-phase ionization refers to the (theoretical) potential at *infinity*, experienced by an electron far away from any matter. In contrast, the energy level relevant for ionization of condensed matter is the *local* vacuum potential, which is associated with a position just outside of a material's surface, where image potentials have diminished to zero but the emitted electrons still experience the extended effects of any condensed-phase interfacial potential.[21] It is the local vacuum level that connects directly to the work function, *i.e.*, the minimum energy required to remove an electron residing at $E_F$ from matter, as shown in Fig. 1. The difference between these two vacuum levels is the residual surface potential, extending beyond the sample surface, denoted as $e\varphi_{outer}$, and associated with, *e.g.*, surface charge or net dipole effects.[21] This residual potential sensitively depends on the specific solution and surface conditions. For neat water, $e\varphi_{outer}$ is expected to be small, between a few to a few tens of meV (see Ref. [9] and references therein; an exact experimental determination of this value remains to be reported). Thus, in the context of neat water, a distinction between these definitions is generally dispensable given the common experimental error bars in LJ-PES experiments. However, if a solute – particularly a surface-active one, as considered here – is introduced or if its concentration is changed, the surface potential may change significantly.

We now introduce the requirements for a successful measurement of a Fermi-referenced PE spectrum from a solution. Such measurements with aqueous solutions are not straightforward and they can only be achieved after favorable experimental conditions have been engineered. Figure 2 gives an overview of the experimental geometry and the various electric-field conditions that can be encountered in LJ-PES experiments. Figure 2A shows a sketch of a typical experimental setup, with the running, *in vacuo* LJ (left) positioned at a small distance, usually <1 mm, from the entrance cone (skimmer) of the detector system (right). Evaporation creates a vapor envelope around the LJ and the intersecting photon beam (green dashed circle) ionizes both phases. In general, a non-vanishing potential difference, which is usually the sum of several potential components, exists between the LJ and the apparatus, even when both are electrically connected to a common ground. The total potential, $V_{tot}$, leads to a different average (de-)acceleration of the photoelectrons originating from the liquid and the gas phase (purple arrows). The acting potentials for this general case of a LJ-PES experiment are detailed further in Figures 2B-2E, where we explicitly show the two components of $V_{tot}$, namely $\Delta e\Phi$ and the undesirable extrinsic potentials. The sign and magnitude of the individual acting potentials is usually unknown and only $V_{tot}$ is revealed *via* a broadening of the gas-phase PE signal and the energy shifts of all PE features.[9] This is exemplified by the PE spectra at the bottom of Fig. 2B-E. A broadened gas-phase PE signal reflects that, on average, the photoelectrons originating from the gas phase experience a potential gradient due to their generation over a range of distances. Hence, they undergo different accelerations, depending on their point of origin, *i.e.*, their distance from the LJ. Note that the liquid-phase PE features are always subject to the full potential, *i.e.*, the gas-phase energy shifts have a somewhat smaller magnitude, on average. Hence, the previously and commonly applied LJ-PES energy referencing to known gas-phase features[18-20, 33] is generally unreliable in the presence of these extrinsic potentials.

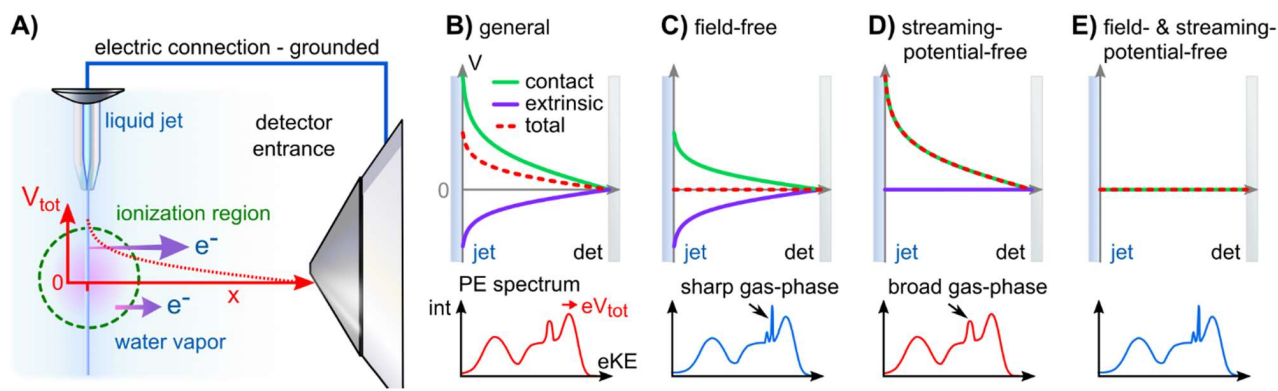

**Figure 2:** Overview of the electric potentials encountered in liquid-jet (LJ) PES experiments. **A)** Sketch of the experimental setup. The LJ (injected from the top) is ionized by the radiation, propagating out of the figure plane (green dashed circle). Photoelectrons are extracted through a differential pumping stage (skimmer to the right). In general, an electric potential (red) exists between the LJ and the detector orifice, which influences photoelectrons (purple arrows) differently depending on their point of origin. (B) Condition for an electrically grounded LJ of an arbitrary solution with a charged surface: charge contributions from the contact potential difference (green) and extrinsic potentials (purple; arising largely from electrokinetic charging or charge-up due to insufficient conductivity). The resulting potential ($V_{tot}$, red dashed curve) corresponds to non-field free / non-streaming-potential free conditions between the LJ and grounded analyzer, causing the gas-phase signal to shift and broaden (see PE spectrum below). In this case, the energy referencing is erroneous. (C) Solution with a precisely tuned salt concentration, exactly compensating the contact potential difference via extrinsic potentials, to achieve a field-free condition for gas-phase referencing; the water gas-phase PE features appear sharp under such conditions. (D) Precisely suppressed extrinsic potential (the condition achieved in the present study). This is the condition required for correct $E_F$-referencing and work-function determination from solution. (E) Rare situation of both contact potential and streaming potential being zero, essentially achieving situations C) and D) at the same time; this is encountered for the TBAI aqueous solutions studied here at a concentration around 20 mM.

Figure 2C illustrates a condition which we term 'field-free', indicating the absence of the total potential between LJ and apparatus. Under this condition, the nascent kinetic energies of the photoelectrons from both the liquid and gas phases can be directly measured. Here, the cumulative extrinsic potentials are tuned by addition of a precise, small amount of solute, which is just sufficient to build up a streaming potential that, in combination with possible potentials from ionization charge-up, counteracts the contact potential difference to an extent that nullifies the total potential.[9, 32, 33] As electrons experience no field gradient, regardless of their origin, the gas-phase signal appears narrow ('sharp') in the spectrum. In aqueous solutions, a handy criterion for judging the sharpness of the H$_2$O 1b$_1$ gas-phase peak is the resolution of its vibrational structure,[35] as shown in Fig. 3 for a 25 mM TBAI$_{(aq)}$ solution. Yet, the sharpness of gas-phase PE peaks also depends on the ionizing light's spot size, which controls the spatial extent of the region around, *i.e.*, distance from, the LJ that is sampled. In the present experiment, the spot size of the implemented plasma-discharge light source is sufficiently large (approximately 300 μm in diameter) to support this approach; see Fig. 2A. In the case of a very small focal spot size, the gas-phase peaks may appear sharp, even in the presence of a field gradient. Under established field-free conditions, liquid-phase PE features have been energy-referenced to the gas-phase PE features; here, peak positions of the simultaneously measured water gas-phase peaks were referenced to well-known IE values to establish the IE energy scale of the liquid phase.[18, 19, 33, 36] Particularly, when a solution has a negligible surface dipole potential – as is the case for nominally neat liquid water – it is possible to infer VIE$_{vac}$ from measurements referenced to the relevant gas-phase vacuum potential at *infinity*, which is essentially equivalent to the *local* vacuum level under such specific conditions. It is important to emphasize, however, that the two vacuum level energies will generally differ for arbitrary aqueous solutions, specifically due to the presence of solution interfacial dipole potentials. This implies that IEs referenced to $E_{vac}^{\infty}$ (gas-phase method) and $E_{vac}^{loc}$ (cutoff method) differ as well; the latter being the relevant reference level for the liquid phase (see Fig. 1). Furthermore, finding the right experimental parameters to achieve field-free conditions is time-consuming and prone to errors. For these reasons, we have developed an alternative, accurate and robust method of measuring VIE$_{vac}$ with respect to the local vacuum level and from arbitrary solutions.[9]

Note that the condition of a suppressed contact-potential difference, presented in Fig. 2C, is not suited to the measurement of Fermi-referenced spectra. Fermi-level alignment, by definition, requires that the vacuum levels of the sample and the detection system are not aligned if the associated work functions are different (which leads to ΔeΦ in the first place; see Fig. 1). Figure 2D shows the condition where only the extrinsic potentials, assumed here to be equivalent to the streaming potential, have been nullified. This *streaming-potential-free* condition is not to be confused with the *field-free* condition (shown in Fig. 2C). The contact potential difference persists exclusively in the former case, making it possible to reference the solution electron energetics to E$_F$, since it can be expected that the internal Fermi level of the solution is energetically aligned with the known Fermi level of the apparatus. (We note that field-free conditions were referred to as 'streaming-potential-compensated' in previous works,[33] before the contribution of ΔeΦ was recognized.[9, 34]) To emphasize, a reference E$_F$ value can only be applied to PE spectra from solution if it can be safely assumed that spectral features have not been shifted by extrinsic potentials. Because ΔeΦ is generally non-zero, and in fact can be greater than one volt in some cases, the gas-phase peak is somewhat broadened and shifted as compared to the field-free case; see the bottom part of Fig. 2D and Ref.[9]. E$_F$ of the apparatus is determined by separately measuring a metallic reference sample. Optionally, the work function of the apparatus may then be determined *via* eΦ$_{det}$ = hν - E$_F$, but not before precisely calibrating the measured energy scale, *e.g.*, by measuring the PE spectrum of a reference gas. Note that if the solution itself is metallic, with a measurable E$_F$, then a reference metal electrode is not needed. Somewhat extreme examples were recently reported by some of us, namely a highly concentrated solution of electrolytes in liquid ammonia[37] and metallic water solution.[38]

Finally, Fig. 2E shows a condition where both the extrinsic potential and ΔeΦ are approximately zero, which implies that the work function of the sample matches that of the apparatus. Energy referencing to both the gas-phase vacuum level and the Fermi level can then be achieved at the same time. However, since the value of $e\Phi_{det}$ depends on the material used for the apparatus, these conditions may only be achieved by coincidence. We will argue later that this happens to be the case for our apparatus and an aqueous solution of ~20 mM TBAI.

Here, we will outline the procedure and conditions for measuring $e\Phi$ and $IE_{EF}$ before applying our approach to aqueous TBAI surfactant solutions. That is, we will disentangle and determine both work function (surface-potential effects) and solute and solvent IE changes (bulk or interfacial electronic-structure effects) in TBAI aqueous solutions as a function of solute concentration. We will argue, based on the analysis of flow-rate-dependent measurements, that extrinsic potentials are negligibly small for $TBAI_{(aq)}$. Using this example, we outline the necessary conditions for measuring Fermi-referenced, liquid-phase spectra and determining work functions. Furthermore, we discuss the challenges of applying this method to an arbitrary solution, where it is difficult to accurately judge remaining extrinsic potentials present in the experiment, most notably the streaming potential. A viable approach would be to quantify, and correct for, streaming-potential contributions during the experiment. This may be achieved by the *in-situ* monitoring of the source of the streaming potentials, namely the streaming current generated at the nozzle orifice. We will discuss such an approach at the end of this paper.

## Experimental Methods

The experiments were performed using the *EASI* (*Electronic structure from Aqueous Solutions and Interfaces*) liquid-jet PES setup[39] which is equipped with a high-energy-resolution, state-of-the-art hemispherical electron analyzer (HEA, Scienta Omicron HiPP-3) and a VUV, monochromatized helium plasma-discharge light source (Scienta Omicron VUV5k). A 40.814 ± 0.002 eV photon energy was implemented in all experiments, as selected by a curved diffraction grating and directed into a 300-μm-inner-diameter glass capillary, yielding a 300 × 300 μm² focal spot size at the sample. The energy resolution of 2 meV was given by the intrinsic width of the selected emission line, He IIα. The monochromatized plasma-discharge source yields an on-target photon flux of $\sim 1\cdot 10^{10}$ photons/s. The emitted light is essentially unpolarized, with the light propagation axis set to an angle of approximately ~110° with respect to the photoelectron detection axis and orthogonally to the LJ axis. The electron detection and LJ axes were set to be orthogonal to each other.

The aqueous solutions were injected using a Shimadzu LC-20 AD high-performance liquid chromatography (HPLC) pump and an inline-degasser unit (Shimadzu DGU-20A$_{5R}$) through a 30-μm-inner-diameter pinhole of a grounded platinum–iridium (PtIr) microplate (Plano GmbH). A flow rate of 1.0 ml/min was implemented, yielding a flowing microjet with ~24 m/s velocity. We have performed comparative measurements with a glass capillary, which is more commonly used in LJ-PES, and confirmed identical results. The advantage of the PtIr microplate, as also used in our previous work,[9, 18, 19] is that the design is more suitable for Fermi-referenced measurements. The metallic microplate allows the direct measurement of the Fermi-level spectrum from the PtIr-solution-injection microplate and provides a more direct grounding of the solution at the point of injection into vacuum (in the case of the glass capillary, the grounding is achieved *via* a metallic tube placed in between the PEEK liquid delivery line, prior to injection into the vacuum chamber). To measure the sample and spectrometer Fermi level, the PtIr-plate mount has a dedicated cutout, through which the plate can be directly exposed to the ionizing radiation.[40] The PtIr target was brought into the light focal spot and aligned towards the analyzer orifice by slightly repositioning the sample rod assembly.

The solution temperature was kept at 10 °C by water-cooling the LJ rod using a chiller unit; evaporation in vacuum leads to associated cooling, and the temperature is expected to be a few degrees lower at the point where the experiments are performed. The vacuum LJ exhibits a laminar-flow region extending over ~5 mm right after the injection point, which subsequently breaks up into droplets due to Rayleigh instabilities.[41] The resulting liquid spray is frozen out and collected at the surface of a liquid nitrogen cold trap, downstream of the injection nozzle and experimental interaction point. Because of water's low conductivity of less than 1 µS/m, even the nominally neat water used in LJ-PES experiments must contain a small amount of electrolyte (usually 5-50 mM) to assure sufficient electrical conductivity and avoid sample charging under irradiation.

The jet's laminar-flow region was positioned at a ~0.8 mm distance from the HEA and ionized in front of its entrance aperture of 0.8 mm diameter. Accurate positioning of the jet was achieved by mounting the LJ assembly on a high-precision x-y-z manipulator. Prior to the measurements, all the surfaces in the vicinity of the LJ-light interaction point were cleaned and graphite-coated to assure a common electric potential and equal work function of all surfaces. Under jet-operation conditions, the average pressure in the interaction chamber was typically maintained at $\sim 2 \times 10^{-4}$ mbar, as accomplished with two turbomolecular pumps (with a total pumping speed of ~2600 L s$^{-1}$ for water) and two liquid-nitrogen cold traps (with a total pumping speed of ~35000 L s$^{-1}$ for water). Aqueous solutions were prepared by dissolving tetra-*n*-butylammonium iodide (TBAI) in highly demineralized water (conductivity ~1 µS/m) and pre-degassed using an ultrasonic bath. For very low TBAI concentrations, NaCl was additionally dissolved to 5 mM concentration to ensure sufficient conductivity and avoid sample charging.

## Results and Discussion

We will firstly present an overview of the experimental PE spectra under investigation and discuss how they relate to the energy scales and intrinsic solution properties we aim to extract. Exemplary PE spectra of reference water (with NaCl salt added to 50 mM concentration to engender electrical conductivity, but with the resulting spectra being otherwise indistinguishable from PE spectra of neat water, blue) and 25 mM TBAI aqueous solution (red) are shown in Fig. 3, together with a separately measured Fermi-edge spectrum (black) from the PtIr microplate. Spectra are presented on an as-measured kinetic-energy scale, eKE$_{meas}$ (bottom axis). With the position of E$_F$ determined via a fit with the Fermi function, as shown in the plot in green, it is straightforward to introduce the IE$_{EF}$ energy scale (A), shown above the spectra, starting with 0 eV defined at E$_F$. VIE$_{EF}$ values for the PE features of interest can directly be determined from the energetic distance from the valence-ionization peak centers to E$_F$. It is important to note that the measured eKE position of E$_F$ in the spectrum does not depend on the sample, as it is a property of the apparatus. In fact, for this work we have measured the Fermi edge repeatedly over the course of months, and a stable value of eKE$_{EF}$ = 36.296 ± 0.008 eV using a 40.814 eV photon energy was confirmed from both the PtIr microplate and a separately mounted gold wire; within the error bars, this value is the same as that reported in our previous study.[9]

The above presentation requires that the conditions for a proper Fermi-level referencing are met, *i.e.*, that the streaming potential, as the main extrinsic potential, is nullified. This is indeed the case, as we will briefly discuss in the following. For the reference water, this is fulfilled at the particular electrolyte concentration and flow speed employed here and consistent with Fig. 2D. Briefly, in Refs. [32, 33], it has been shown that the streaming current, I$_{str}$, which is the source of $\Phi_{str}$, crosses zero at this specific electrolyte concentration at a flow-speed regime of ~17-30 m/s (*e.g.*, a flow rate of 0.5 ml/min with a nozzle orifice diameter of 25 µm, as used in Refs. [32, 33]), similar to the ones implemented in our experiment; also see Ref. [9]. Thus, $\Phi_{str}$ is suppressed in the liquid jet used to produce this reference spectrum. Note that the interaction of the solution with the walls forming the orifice channel of the PtIr microplate or the inner walls of the glass nozzle used in these previous studies can in principle be different. However, our glass-capillary-based cross-check experiments reveal the same results, highlighting a negligible streaming current in both cases. The PtIr microplate is preferred in the present study, as detailed in the experimental section and the SI.

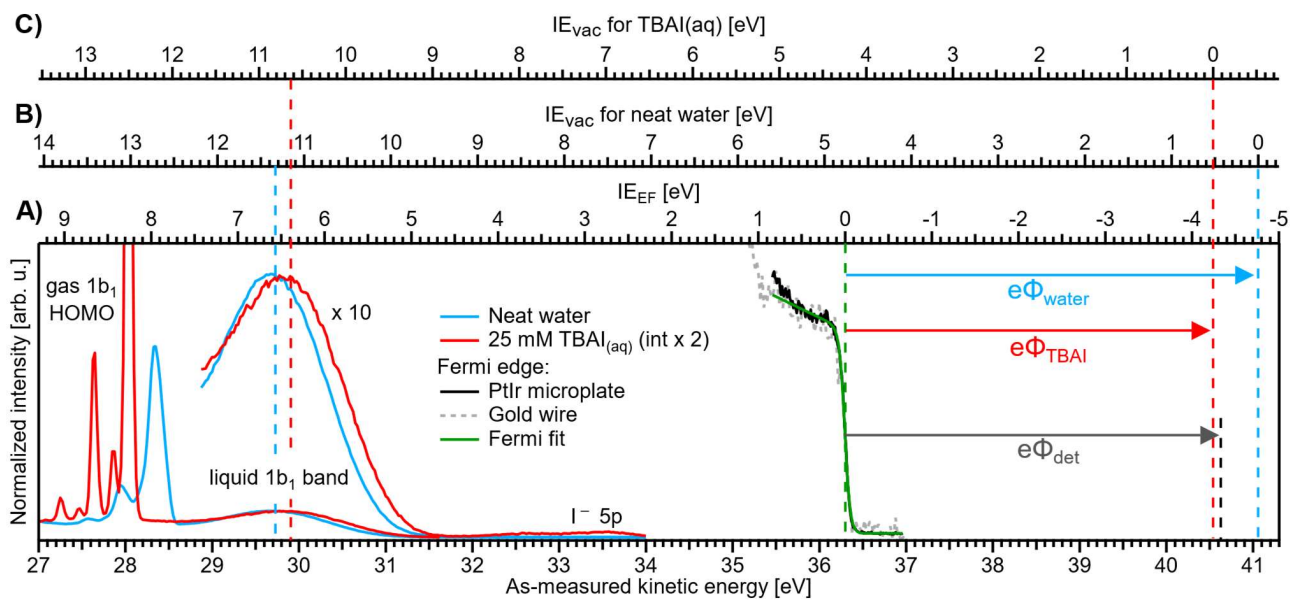

**Figure 3:** Valence PE spectra of reference liquid water (blue) and 25 mM TBAI$_{(aq)}$ (red), measured from a grounded liquid jet and at hν = 40.814 eV; the bottom axis shows the as-measured eKE scale of the detector. Measurements were performed under conditions free of any potential other than the contact potential difference, ΔeΦ (compare Fig. 2D). The Fermi-edge spectrum (black) was recorded from the PtIr microplate and fitted with a Fermi function (green line), where the center position defines the zero point of the IE$_{EF}$ energy scale, **A)**. For reference, a Fermi-edge spectrum was measured from a gold wire under similar conditions and is shown in grey. The IE$_{vac}$ axes for reference water, **B)**, and 25 mM TBAI$_{(aq)}$, **C)**, are defined using the liquid-water 1b$_1$ peak as an anchor point (compare the red and blue dashed lines, respectively). The difference between VIE$_{EF}$ and VIE$_{vac}$ for each case gives, per definition, the solution work function, eΦ (arrows to the right). The apparatus work function, eΦ$_{det}$ (black arrow, after energy-scale correction) is added for comparison and has almost the same value as that of the 25 mM solution.

For the TBAI$_{(aq)}$ solution, we confirmed the absence of Φ$_{str}$ via auxiliary measurements of PE spectra as a function of flow rate over a wide range (0.5-2.0 mL/min.) for two different TBAI concentrations, 12.5 mM and 25 mM TBAI$_{(aq)}$; see the SI for details. As Φ$_{str}$ depends on the flow rate, any non-vanishing Φ$_{str}$ should manifest in a change in the observed PE peak positions when the flow rate is altered. However, for both concentrations, the peak position of the liquid 1b$_1$ feature changed by no more than ~15 meV when extrapolated to zero flow rate, where Φ$_{str}$ must vanish. This value is well below our experimental IE-determination accuracy of ~30 meV, and the streaming potential can be correspondingly neglected. As to why this value is so small in the case for TBAI solutions, we can currently only speculate. The streaming current, I$_{str}$, is generated by a charge separation at the inner walls of the capillary used to form the liquid jet.[19, 28] Here, ions strongly interact with the capillary wall and form the inner Helmholtz layer.[31] Incidentally, the flow rate is also zero close to the wall. Counterions compensate the accumulated charge imbalance in the outer mobile Helmholtz layer, flowing with the liquid stream. As TBA$^+$ is particularly hydrophobic, it can be assumed that the inner Helmholtz layer consists of these molecules. One could imagine that the carbon chains effectively screen or physically distance the rest of the solution from the capillary wall, which may mitigate the solution–wall interaction and thus the charge accumulation in the outer Helmholtz layer. Another possibility may be that the large physical size of TBA$^+$ makes it more prone to be dragged along with the liquid stream, reducing the electrokinetic charge separation.

Ionization energies have previously been determined with respect to the *local* vacuum level, IE$_{vac}$, for both neat water and TBAI aqueous solutions via the application of a negative bias voltage to the liquid jet.[9, 15] We can correspondingly translate the $IE_{vac}$ energy scale to the data from the grounded liquid jet by referring to an easily identifiable PE feature, such as the liquid-water $1b_1$ band, and fixing the energy scale to this peak using the $VIE_{vac,1b1}$ position for each solution (see the dashed lines to the left in Fig. 3). In principle, $IE_{vac}$ values can be determined by simply measuring the biased PE spectra of the low-energy tail and the PE feature of interest, as described in our previous work.[9] Here, due to the time-consuming nature of measuring both biased and grounded spectra over a large range of solute concentrations, we utilize the results from our previous works, $VIE_{vac,1b1}$ = 11.33 eV for neat water[9] and $VIE_{vac,1b1}$ = 10.63 eV for 25 mM $TBAI_{(aq)}$ (also denoted as 1 ML TBAI in some of our previous works).[9, 15] This yields the $IE_{vac}$ energy scales (B) and (C) in Fig. 3 for the reference water and TBAI aqueous solution, respectively. We can now combine both results to determine each solution's work function via $e\Phi = VIE_{vac} - VIE_{EF}$ (compare arrows at the right of the figure). This figure illustrates the TBAI-induced energetic changes of the water VIE values, as discussed below. We note that this procedure was already tentatively applied to PE spectra from neat water and an exemplary solution of 25 mM $TBAI_{(aq)}$ in our previous study[9], and the values are found to be in excellent agreement with those reported here. In Fig. 3, we also present $e\Phi_{det}$ = 4.30 ± 0.04 eV (black dashed line) for comparison (to calculate $e\Phi_{det} = h\nu - eKE_{EF}$ we determined $E_F$ on a calibrated kinetic-energy scale; see the SI for details).

We next turn to the quantitative energy changes of water's valence PE features, with a focus on the $1b_1$ HOMO band, as obtained under a systematic variation of TBAI concentration. Importantly, we found that solutions with TBAI concentrations of less than 2 mM are insufficiently conductive to reliably perform LJ-PES measurements. This necessitated the addition of a small amount of salt; as mentioned in the Experimental Section we used 5 mM NaCl, which is sufficient to measure PE spectra from liquid water.[9] With an effective segregation leading to a bulk-to-surface enrichment of approximately 70:1,[22] even 0.1 mM $TBAI_{(aq)}$, the lowest concentration employed here, corresponds to an approximately 7 mM concentration of interfacial iodide. This is larger than the 5 mM NaCl bulk solute concentration, especially when considering that $Cl^-$ is less likely to accumulate at the interface, which is probed by PES.[41] We thus consider the added NaCl to be, at most, a small perturbation to the studied system. The results, shown later, indicate that there is a gradual transition of spectral changes with TBAI concentration, unaffected by the inclusion of NaCl salt, as long as the solution conductivity is sufficiently high. That is, there is no sudden spectral shift when switching between reference water (containing 50 mM NaCl) and 0.1 mM TBAI + 5 mM NaCl aqueous solution, or between 0.5 mM TBAI + 5 mM NaCl aqueous solution and 1 mM $TBAI_{(aq)}$ without any NaCl.

A series of measured valence PE spectra, focusing on the lowest ionization energy, $1b_1$ PE features of water and covering the 0-40 mM TBAI concentration range, are presented in Fig. 4. The $TBA^+$ and $I^-$ (labelled $I^-$ 5p in Fig. 3) spectral features have been discussed in our previous study, where we also explored multiple TBAI concentrations in a biased configuration (implying that the associated work functions could not be directly accessed).[15] Panels A and B of Fig. 4 respectively show the gas-phase and liquid-phase spectral regions on separate scales, as their peak intensities largely differ (see Fig. 3). The spectral intensities have been normalized to the water $1b_1$ HOMO peak height, with the v = 0 vibrational peak of the gas-phase signal occurring between ~27.9 and 28.3 eV and the wide, liquid-phase $1b_1$ band appearing near 29.7 eV, on the as-measured electron kinetic-energy scale, $eKE_{meas}$. Note that the broader widths of the liquid-phase peaks largely arise from the broad distribution of hydration configurations in liquid water.[18] The data shown without the normalization and intensity offsets can be found in Fig. S3 of the SI. The bottom spectrum in Fig. 4 is measured from reference water, with 50 mM NaCl and containing no TBAI; the same spectrum is shown in Fig. 3. We use these conditions as a streaming-potential-free starting point for our grounded-solution TBAI-concentration series.

We remind the reader that the gas-phase PE shifts follow the evolution of $V_{tot}$, while the peak width reveals $|V_{tot}|$. $V_{tot}$ is equal to $\Delta e\Phi$ in the present case due to the absence of $\Phi_{str}$. $\Delta e\Phi$ in turn depends on the work function

of the apparatus, and thus depends on the experimental apparatus in use. If the gaseous peak width is small, this means that $V_{tot}$ crosses zero. The liquid-phase PE features instead shift according to changes in $IE_{EF}$. The shifts shown in panels A and B of Figure 4 thus reveal different aspects of the concentration-dependent changes, which happen to occur in the opposite direction in the present case.

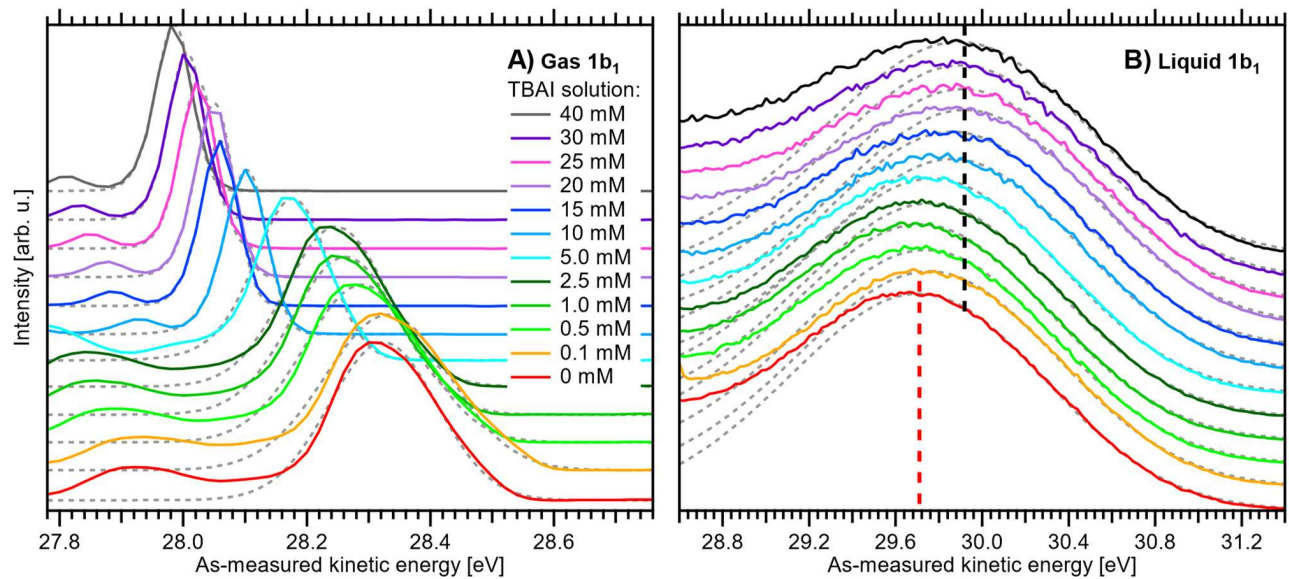

**Figure 4:** Experimental PE spectra of TBAI$_{(aq)}$ as a function of concentration from 0 to 40 mM: **A)** Close-up on the v = 0 vibrational peak of the gas-phase 1b$_1$ (HOMO) water orbital ionization feature and **B)** close-up on the liquid-phase 1b$_1$ HOMO ionization band, each normalized to the peak maximum and vertically offset; both features were measured together but are shown separately for clarity. For measurements with TBAI concentrations ≤1 mM, an additional 5 mM concentration of NaCl salt was dissolved to maintain the electrical conductivity of the solution. Fitted peak functions are added as dotted grey lines. In panel B, dashed lines indicate the reference water 1b$_1$ (Gaussian) peak position (red) and its most shifted peak position, associated with the 40 mM TBAI$_{(aq)}$ solution (black). The same data are plotted without normalization and offsets in Fig. S3 of the SI. See the main text for details.

All spectra in Fig. 4 were fit with two Gaussian functions, one for the dominant (v = 0) gaseous and one for the liquid 1b$_1$ peak, which are shown as grey dotted lines. Because of overlapping features from both phases, it is not possible to accurately fit the full spectrum; such analysis may be performed for a biased measurement, essentially in the absence of the gas-phase signal.[9, 15] Instead, we chose to fit only the PE feature of interest with an exponentially modified Gaussian (EMG) shape,[42] which describes the intrinsic peak asymmetry well. This is straightforward for the gaseous peak, which is well isolated; note that in the broadest case, the v = 0 peak overlaps with the $v_{bend}$ = 1 (symmetric bend) vibrational peak,[35] which is small enough that it does not affect the result. The liquid-phase band, however, significantly overlaps with the 3a$_1$ (HOMO-1) photoelectron band and, at higher TBAI concentrations, TBA$^+$ signal contributions.[15] To avoid skewing the water 1b$_1$ component by these contributions, the fit was performed using the same fixed asymmetry, τ, value of -0.3 as in our previous work[15] and limiting the fit range to the high eKE shoulder of the band. The fitted peak seems to lean towards the higher-eKE side of the band at higher TBAI concentrations (see Fig. 4B). This is the expected behavior, as seen in the full fits to the liquid-only spectra reported in our previous study.[15]

We now evaluate the quantitative evolution of the 1b$_1$ liquid-phase peak position retrieved from peak fits shown in Fig. 4B; we will discuss the corresponding evolution of the gaseous peak (Fig. 4A) later. As shown in Fig. 5B, from the smallest to the highest TBAI concentration, 0 mM to 40 mM TBAI, the peak position (as-measured eKEs, left axis) increases slightly and gradually, by a total of approximately 190 meV, from 29.71 to 29.90 eV KE, where the shift appears to saturate. There may be an indication of a small shift back towards lower eKEs at very high concentrations, but this is within the experimental error. Error bars originate from a quadratic addition of fit errors and the confidence interval determined from repeated measurements. Analogous to Fig. 3, we introduce the $IE_{EF}$ energy scale, shown on the right side of Fig. 5B and established *via* referencing to the Fermi level of the apparatus. We recall that this reference is only correct under the assumption that any extrinsic potential is absent, as has been established in the context of Fig. 3, and note again that this is an exceptional situation, and that other solutions and different nozzle configurations might have a considerably larger contribution from extrinsic potentials, preventing Fermi referencing without correction efforts. On the $E_F$ scale, the $1b_1$ peak position starts at $VIE_{EF} = 6.58 \pm 0.03$ eV for reference water and reaches $VIE_{EF} = 6.39 \pm 0.03$ eV at 30 mM $TBAI_{(aq)}$, in excellent agreement with the tentative results reported in our previous work.[9] The liquid-phase peak widths increase by approximately 30%, again in agreement with our previous study using a bias voltage (see Fig. S4 in the SI).[15] Hence, the observed broadening is not caused by the varying potentials, but is rather an intrinsic solution property.

We are now set to determine $e\Phi_{liq}$ as a function of TBAI concentration. In principle, it would suffice to simply subtract the $VIE_{EF}$ values presented in Fig. 5B from the $VIE_{vac}$ values determined in our previous study (reproduced in Fig. 5C),[15] *i.e.*, applying $e\Phi = VIE_{vac} - VIE_{EF}$. However, to avoid any influence of the different fitting methods employed to extract the VIE values, we instead opt to directly match the experimental PE spectra referenced to their respective energy scales, *i.e.*, the Fermi-referenced PE spectra from our grounded measurements in this work *versus* the vacuum-referenced PE spectra from the biased measurements reported in Ref.[15]. For this, a fit is employed that energy-shifts the grounded, Fermi-referenced PE spectrum onto the biased, cutoff-referenced PE spectrum of Ref.[15] to find the optimal overlap of the water solvent $1b_1$ band. An exemplary fit for the spectra from 25 mM $TBAI_{(aq)}$ is shown in Fig. 6A. PE contributions from ~0.6 eV below the peak maximum were excluded from the fit (purple dashed curve in Fig. 6A). The resulting shift is, per definition, equal to $e\Phi$ of the solution. The results for all concentrations are plotted in Fig. 6B. The values for reference water, $e\Phi_{water} = 4.76 \pm 0.03$ eV (blue dashed line), and the apparatus, $e\Phi_{det} = 4.30 \pm 0.04$ eV (red dashed line, as introduced earlier), are indicated as well. We have previously estimated the contact potential difference between liquid water and our apparatus, under nominally identical experimental conditions as in the present study, to be approximately 0.43 V.[9]

Upon addition of small amounts of solute to water, starting from 0.1 mM TBAI, $e\Phi_{liq}$ decreases rapidly. A clear deviation from this trend is observed above ~10 mM, when $e\Phi_{liq}$ decreases more slowly; we found saturation of such effects starting between 10-15 mM in our previous work.[15] Notably, $e\Phi_{liq}$ crosses $e\Phi_{det}$ of the apparatus in the 15-25 mM TBAI concentration range, which leads to the field-free *and* streaming-potential-free conditions discussed above, related to Fig. 2E. Indeed, the gaseous $1b_1$ peak width (Fig. 4A and Fig. S5 in the SI), which is an indication of the absolute acting potential $|V_{tot}|$, shows a minimum in this concentration range, indicative of the absence of broadening from a potential gradient between the LJ and the analyzer, *i.e.*, confirming that indeed $V_{tot} \approx 0$ V.

We briefly come back to the gaseous peak positions shown in Fig. 5A, which show a striking resemblance to the $e\Phi_{liq}$ curve in Fig. 6B. This is no coincidence, as we observe here the influence of $V_{tot} = \Delta e\Phi$ on the gas-phase PE features: the gas-peak position shifts towards lower KE according to the total acting potential, $V_{tot}$. This becomes even more apparent when overlaying the averaged $1b_1$ gas-phase energy shifts of Fig. 5A with the observed work-function change in Fig. 6B; see the dashed purple line. Note that the gas-phase peaks experience less than the full applied potential, *i.e.*, $cV_{tot}$, with the geometry correction factor, $c$, describing the geometric details of the particular setup, such as sample-to-skimmer distance and irradiated spot size.[43] An almost perfect match was achieved when the gas-phase shift was scaled by a factor of 1.5. Thus, we find $c = 1/1.5 \approx 0.67$ for our experimental setup.

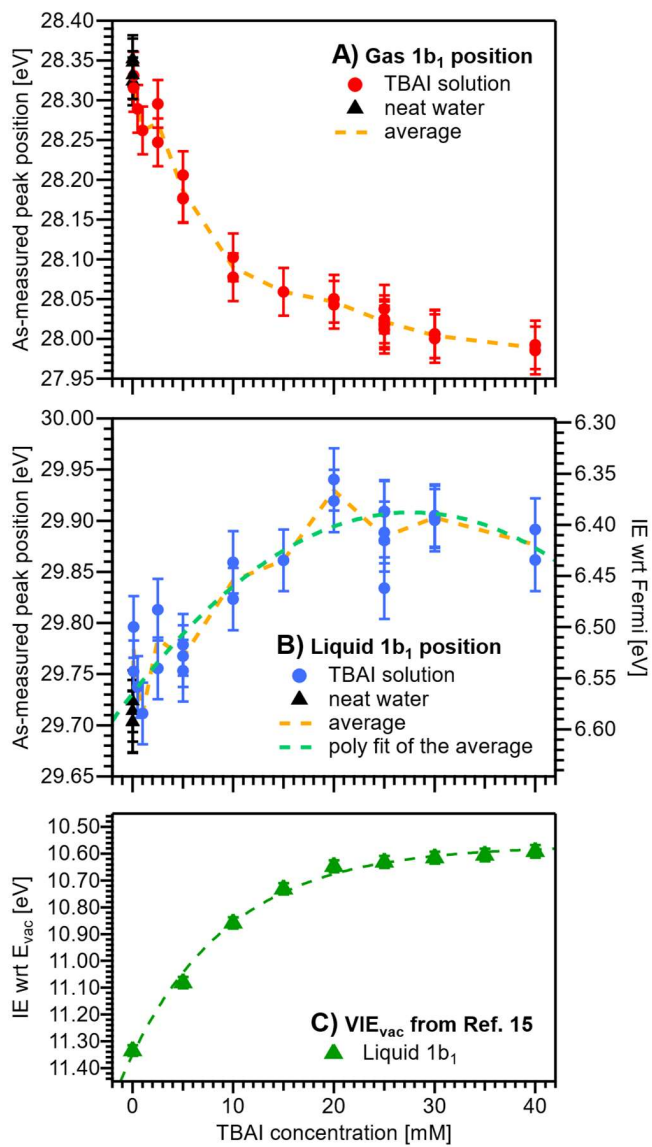

**Figure 5:** Position of the gaseous **A)** and liquid **B)** $1b_1$ peaks extracted from fits to the PE spectra as a function of TBAI concentration. Red A) and blue B) dots indicate results from TBAI$_{(aq)}$ (5 mM NaCl have been added for concentrations at and below 1 mM TBAI) and black triangles indicate reference water (no TBAI but 50 mM NaCl added). In panels A and B, the left axis shows the as-measured eKEs. In panel B, the right axis additionally shows the IE$_{EF}$ scale, *i.e.*, peak VIE values referenced to the Fermi level; this scale assumes streaming-potential-free conditions throughout (see Figs. 2D and 2E). The error bars represent the experimental reproducibility in determining IE values (30 meV). The standard deviation extracted when averaging multiple measurement values at each concentration was 30-40 meV (not shown in the figure for clarity). **C)** Cutoff-calibrated IE$_{vac}$ values for the water $1b_1$ peak as function of TBAI concentration, adapted from Ref. [15], which represents VIE$_{vac}$ = e$\Phi$ + VIE$_{EF}$. The green dashed line is a fit to the experimental data. Note that the change in IE$_{EF}$ is only a fraction of the change in IE$_{vac}$, because the latter is mainly driven by changes in e$\Phi$.

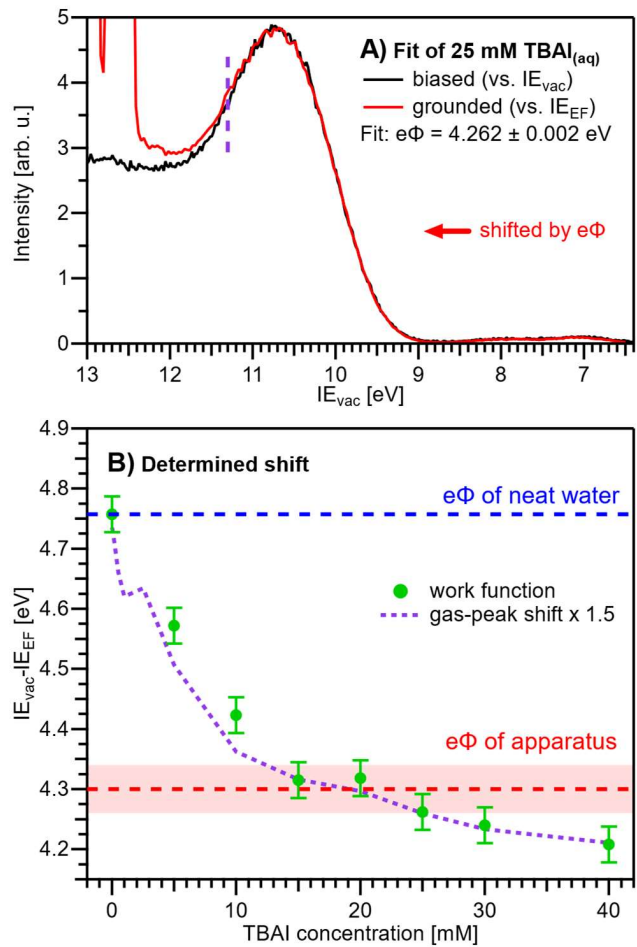

**Figure 6: A)** Exemplary fit of a (grounded) $IE_{EF}$-referenced PE spectrum, here for 25 mM $TBAI_{(aq)}$, to the (biased) $IE_{vac}$-referenced spectrum presented in our previous work.[15] The fit finds the optimal overlap of the $1b_1$ peak by shifting the grounded spectrum by a specific amount, where, per definition, the shift value corresponds to the work function, $e\Phi$. Only PE spectral contributions to the right of the purple dashed line were considered in the fit. **B)** $e\Phi_{liq}$ values resulting from fitting the experimental PE spectra at each concentration (green dots). For this procedure to work, it must be ensured that the only acting potential is $\Delta e\Phi$ (see Fig. 2D). The apparatus work-function value, $e\Phi_{det}$, is indicated with a red dashed line. $e\Phi_{liq}$ crosses $e\Phi_{det}$ between 12-25 mM, *i.e.*, here $V_{tot} \approx 0$ V (also see Fig. S5). The averages of the values from Fig. 5A (gas-peak shift) are added as a purple dotted line for comparison, here multiplied by an empirical factor of 1.5 and shifted to match the reference water value (zero TBAI concentration). Error bars indicate our experimental reproducibility.

We finally bring both the $VIE_{EF}$ and $e\Phi_{liq}$ data together into a unified picture, enabling us to directly quantify the competing contributions of work function and electronic structure to the previously measured concentration-dependent $IE_{vac}$ changes. This is presented in Fig. 7, with the work-function data in green from Fig. 6B, $VIE_{EF}$ data in blue from Fig. 5B, and the vertical arrows representing the $VIE_{vac}$ values at two exemplary TBAI concentrations, reference water and 40 mM TBAI (compare Fig. 5C). These results are also summarized in Table 1. Referring to both Table 1 and Figure 7, we make several important and rather unexpected observations. First, the work function rapidly decreases, nearly linearly, by almost 0.4 eV when going from reference water to ~10 mM TBAI; for higher concentrations the decrease is much smaller, with only a ~100 meV change in the range of 10-40 mM TBAI. The second crucial finding is that the work-function effect (in total ~520 meV) accounts for approximately ~75% of the (previously determined)[15] changes of $IE_{vac}$, with ~25% being due to solute-induced water-solvent electronic-structure changes, a conclusion that was impossible to draw in the earlier studies. Qualitatively, the work-function plateau at the higher concentrations can be attributed to the saturated TBAI surfactant monolayer, with possible accommodation of additional solute in a deeper (sub-)surface layer at the highest concentrations. This indeed mirrors the previously reported adsorption behavior, with saturation tending to set in near 15 mM.[15] The change of work function at lower concentrations would be qualitatively assigned to the subsequent buildup of molecular surface dipoles, with a considerable component perpendicular to the solution surface. The decrease of the work function implies a negative surface dipole, associated with a dipole layer with negative charge pointing into the solution and positive charge residing at the top surface. This corresponds to the commonly assumed structure of the $TBA^+$ / $I^-$ segregation layer,[12, 44, 45] with $TBA^+$ ions residing at the very top of the surface and the associated iodide counter ions tending to be located at a somewhat larger distance from the surface. Notably, an earlier theoretical study suggested that no dipole is formed by $TBA^+$ and $I^-$ ion pairs perpendicular to the surface because both cations and anions exhibit strong surfactant activity and thus fail to form a strong electric double layer.[22] This prediction is not supported by the present study. Regarding experimental access to surfactant structure, we mention the application of neutral impact collision ion scattering spectroscopy (NICISS) in combination with Kelvin probe measurements, including measurements of different tetrabutyl-X (X = phosphonium, ammonium) salts in formamide.[46] Those studies revealed different depth profiles of the surfactant molecules and their counter anions, correlating with an electrically charged double layer and a corresponding electric potential at the surface of the salt solutions. We note, however, that such experiments would be much more challenging for aqueous solutions, which have two orders of magnitude higher vapor pressures.

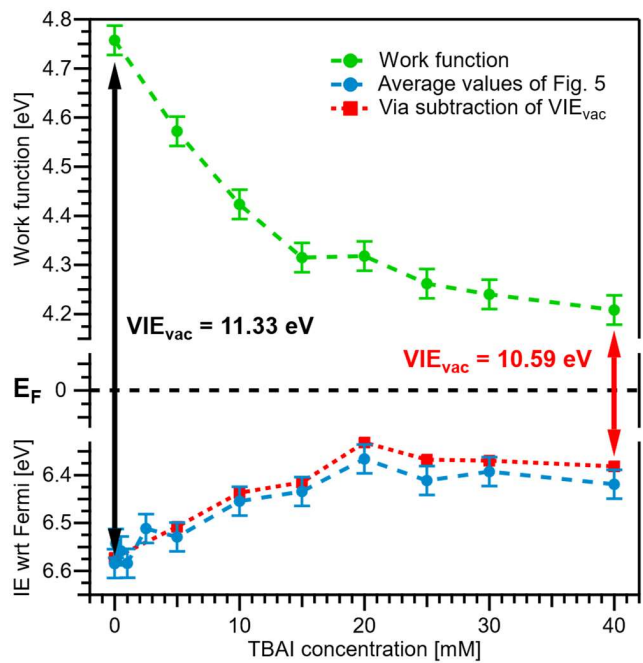

**Figure 7:** Bringing the results from Figs. 5 and 6 together. **Bottom:** Average values from Fig. 5B in blue. **Top:** $e\Phi_{liq}$ results from Fig. 6B in green. The constant Fermi level, $E_F$, in the center is only valid if the solely acting potential is $\Delta e\Phi$ (see Fig. 2D). The distance between the vacuum level, $E_{vac}$, indicated by the work function, and the peak of interest on the Fermi-referenced energy scale, $VIE_{EF}$, must be equal to $VIE_{vac}$, per definition, and this is fulfilled over the whole range. $VIE_{vac}$ values are indicated by the arrows for reference water (black arrow), $VIE_{vac}$ = 11.33 eV,[9] and 40 mM $TBAI_{(aq)}$ (red arrow), $VIE_{vac}$ = 10.59 eV.[15] To confirm the correctness of our analysis, the red dashed line represents an alternative way to determine VIE$_{EF}$: Subtraction of the solute-concentration-dependent VIE$_{vac}$ values from Ref. [15] from the e$\Phi$ values reported here (green) effectively utilizes the fits from the previous study to determine VIE$_{EF}$. The red and blue curves are essentially equivalent, validating the performance of the fitting procedure implemented in Fig. 4B.

| conc. | VIE$_{EF,1b1}$ (eV) | e$\Phi$ (eV) | VIE$_{EF,1b1}$ + e$\Phi$ (eV) | VIE$_{vac}$ (eV) from Ref. 15 |
|---|---|---|---|---|
| 0 mM | 6.58 ± 0.03 | 4.76 ± 0.03 | 11.34 ± 0.03 | 11.33 ± 0.02 |
| 5 mM | 6.53 ± 0.03 | 4.57 ± 0.03 | 11.10 ± 0.03 | 11.08 ± 0.02 |
| 10 mM | 6.45 ± 0.03 | 4.42 ± 0.03 | 10.88 ± 0.03 | 10.86 ± 0.02 |
| 15 mM | 6.43 ± 0.03 | 4.32 ± 0.03 | 10.75 ± 0.03 | 10.73 ± 0.02 |
| 20 mM | 6.37 ± 0.03 | 4.32 ± 0.03 | 10.68 ± 0.03 | 10.65 ± 0.02 |
| 25 mM | 6.41 ± 0.03 | 4.26 ± 0.03 | 10.67 ± 0.03 | 10.63 ± 0.02 |
| 30 mM | 6.39 ± 0.03 | 4.24 ± 0.03 | 10.63 ± 0.03 | 10.61 ± 0.02 |
| 40 mM | 6.42 ± 0.03 | 4.21 ± 0.03 | 10.63 ± 0.03 | 10.59 ± 0.02 |

**Table 1:** VIE$_{EF}$ and e$\Phi_{liq}$ values extracted from fits to the spectra of solutions with various TBAI concentrations and displayed in Fig. 7. The presented errors are a combination of fit errors and the reproducibility of repeated measurements.

The aforementioned insights on the segregation layer structure are, however, too simple to explain the observed plateau of the work function above the saturation concentrations. Our PE experiments suggest that the dipole orientation varies insignificantly until the monolayer is completed unless dipole-compensation effects occur due to the second surfactant layer. Clearly, a satisfying structural model of the TBAI aqueous surface that considers the concentration-dependent TBAI butyl-chain orientation and resulting complex interplay between steric- and electronic-structure effects is lacking. We note that theoretical modeling has been performed only for the highest concentrations,[12, 22] and the calculated surface potential for a 49 mM TBAI aqueous solution is approximately 0.7 eV,[12] *i.e.*, significantly larger than that measured in the experiment.

What can be learned from the smaller, <200 meV, and smoother water 1b$_1$ energy shift associated with electronic-structure changes, as identified in the bottom part of Fig. 7? The 1b$_1$ VIE$_{EF}$ follows the same Langmuir adsorption trend as observed in our previous work[15] and can be explained as the concentration-dependent buildup of an interfacial ion concentration and the associated solute-induced changes of the interfacial water electronic structure. There is no theoretical work to detail this for TBAI aqueous solutions. However, we have previously discussed analogous effects for medium to highly concentrated NaI aqueous solutions, based on experiment and computations.[14] In that case, interfacial ion segregation played a minor role, and the observed small binding-energy shifts can be majorly attributed to ion-induced changes of the water electronic structure. Particularly, when specific hydrating water units are replaced by ions, the intermolecular bonding interaction between the water molecules is weakened, resulting in differential peak shifts of the water orbitals, as well as a notable narrowing of the 3a$_1$ PE feature;[14] the latter being difficult to detect in the TBAI$_{(aq)}$ spectra due to the 3a$_1$ overlap with the TBA$^+$ signal contribution.[15]

For completeness, in the bottom panel of Fig. 7, we also show IE$_{EF}$ values as red squares, which are based on a simple subtraction of e$\Phi_{liq}$ values from the VIE$_{vac}$ values reported in Ref. [15] (see Fig. 5C); in effect, this yields a result as if the full valence-band fit of the previous study would have been applied to the current data. The results agree well with those employing the fitting method presented in Fig. 4B. In principle, the method of fitting biased, cutoff-referenced PE spectra to unbiased, Fermi-referenced PE spectra can be used to easily determine the VIE$_{EF}$ values for other water orbitals or solute PE features. For example, using the biased and unbiased spectra of reference water, one obtains VIE$_{EF,3a1L}$ = 8.36 ± 0.03 eV and VIE$_{EF,3a1H}$ = 9.76 ± 0.03 eV for the split 3a$_1$ orbital of water, *i.e.*, here the obtained work function, shown in Table 1, was simply subtracted from the values obtained in our previous study (analogous to 11.34 ± 0.03 eV = 6.58 ± 0.03 eV + 4.76 ± 0.03 eV).[15] Similarly, for the I⁻ solute feature associated with the 40 mM TBAI$_{(aq)}$ solution we obtain vertical detachment energies (VDEs)[N1] of VDE$_{EF,I5p1/2}$ = 3.8 ± 0.1 eV and VDE$_{EF,I5p3/2}$ = 2.8 ± 0.1 eV. Note that this alternative method requires the measurement of PE spectra under both biased and grounded, streaming-potential-free conditions, which are then referred to their respective energy scales (IE$_{vac}$ and IE$_{EF}$, respectively).

As an outlook, we briefly discuss how the streaming potential, $\Phi_{str}$, could be monitored *in situ*, and possibly corrected for. Such an approach could be implemented in future Fermi-referenced PE experiments of arbitrary solutions, where $\Phi_{str}$ is likely non-zero and a large source of error, unlike in the TBAI$_{(aq)}$ case presented above. The associated streaming current, $I_{str}$, can in principle be measured under the same experimental conditions as adopted in the LJ-PES experiments, and possibly even *in situ* while measuring the PE spectra. The relation between $\Phi_{str}$ and $I_{str}$ for a cylindrical LJ is:[19, 28, 32]

$$\Phi_{str} = -\frac{1}{2\pi\epsilon_0} \frac{I_{str}}{v_{str}} \ln\left(\frac{d}{2R}\right)$$

where $\epsilon_0$ is the vacuum permittivity, $v_{str}$ is the stream velocity, $d$ is the jet diameter, and $R$ is the distance where the potential assumes a zero value (this can be, *e.g.*, the grounded analyzer skimmer cone in our case). Here, the geometrical dimensions of the apparatus are known, and the flow velocity can be determined using the (calibrated) HPLC pump settings and nozzle geometry. Thus, if $I_{str}$ could be measured, a direct calculation of $\Phi_{str}$ would be possible. Faubel *et al.* found an excellent match between the $\Phi_{str}$ value extracted from $I_{str}$ measurements from neat water at the LJ nozzle tip and associated shifts of PE spectra, with $\Phi_{str}$ values being on the order of tens of V.[19] In the work of Preissler *et al.*, correlations of measured $I_{str}$ values and observed PE spectral shifts from an NaI aqueous solution were inconclusive.[32] This was likely due to the low (UV) ionizing photon energy employed in the study, which prevented an accurate IE evaluation due to significantly enhanced (vibrational inelastic) electron scattering in this ionization and electron kinetic-energy regime. As we have reported recently, aqueous-phase PE features with eKEs below ~15 eV are strongly distorted by electron scattering.[47] However, higher photon- and kinetic-energy measurements can be expected to yield more robust correlations. With this in mind, we are currently preparing *in situ* LJ $I_{str}$ measurements from an isolated PtIr microplate that will also be used to record valence PE spectra with a 40.814 eV photon energy, yielding sufficiently high eKEs to avoid deleterious vibrational inelastic scattering effects in the aqueous phase.[47] However, it remains to be seen whether $\Phi_{str}$ corrections on the order of a few 100 meV are feasible from *in situ* $I_{str}$ measurements; the associated currents may only be on the order of 0.1 nA and less, which may be difficult to measure reproducibly.

## Conclusions

LJ-PES has matured into a technique capable of accessing explicit solution-surface descriptors, which was previously limited to solid-phase condensed matter. Specifically, the measurement of work functions, $e\Phi$, from volatile liquids, and particularly from liquid water and aqueous solutions, is now possible under certain conditions, and in principle connects electron energetics with electrochemistry. We have presented an explorative protocol for the quantitative determination of concentration-dependent, Fermi-referenced IEs and $e\Phi$s, and applied it to the representative TBAI surfactant in aqueous solution. Such solution-phase measurements are generally difficult due to the deleterious effects of various, typically hardly quantifiable and competing solution surface-charging mechanisms. This complicates the extraction of the surface potential and the associated $e\Phi$ of interest. Measuring Fermi-referenced PE spectra from solutions requires good electric contact between the analyzer and the solution (to warrant Fermi-level alignment) and that all extrinsic potentials, which would arbitrarily shift liquid-phase PE spectra, are absent during the measurements. The challenge is to quantify extrinsic potentials and, particularly, distinguish them from the contact potential difference, $\Delta e\Phi$, which arises from the difference in $e\Phi$ between the solution and the apparatus. The LJ-PES experiment only allows the sum of all acting potentials, *i.e.*, the total acting potential, $V_{tot}$, to be assessed, but compensation of $V_{tot}$, including $\Delta e\Phi$, implies that the Fermi level is not aligned between the sample, reference metal, and detector and that Fermi-level energy referencing and sample $e\Phi$ determinations cannot be achieved.

Aqueous-phase TBAI appears to be a fortunate system, with the concentration-dependent changes in surface potentials and thus $e\Phi$s being large in comparison to the extrinsic potentials, which have been found to be largely absent under the employed experimental conditions. Specifically, the latter was confirmed by the absence of $\Phi_{str}$ *via* the analysis of energy shifts of the water PE features as a function of flow rate and solute concentration. This is an essential step, since streaming potentials may be large depending on the employed solution. We do not see any indication that extrinsic potentials are significantly above ~15 mV at all TBAI concentrations, enabling us to refer the liquid-phase PE features to the Fermi level. This condition cannot be fulfilled for solutions with, for example, a significant streaming potential, such as even simple electrolyte solutions. As a possible future development, *in situ* streaming current measurements are proposed, which may allow numeric compensation of streaming-potential contributions and thus Fermi referencing of arbitrary solutions.

The remaining step to generally determine Fermi-referenced IEs is the assignment of the Fermi energy to the solution spectra, which then allows the solvent and solute energetic positions to be defined on the $IE_{EF}$ energy scale. To determine $e\Phi$, a second quantity must be measured for the same system, namely the IE relative to the vacuum level, $IE_{vac}$. The work function is then calculated as $e\Phi = IE_{vac} - IE_{EF}$. $IE_{vac}$ values can be accurately measured as the energy difference of the low-energy cutoff of the solution photoelectron spectrum and the detected eKE of a solvent or solute feature of interest under biased conditions, provided the photon energy is exactly known. This accurate energy-referencing method, which we previously introduced,[9] has been a key development in determining solution $e\Phi$s. Proper determination of the other quantity, $IE_{EF}$, has been explored here, and involves the careful preparation of experimental conditions without extrinsic potentials. We found that an optimal way of determining exact $e\Phi$ values is to measure both the Fermi-referenced and biased-sample PE spectra with the same system under otherwise similar conditions and then directly fit the spectral shapes of both to obtain their energetic separation, and thus $e\Phi$, with minimal error.

Now, combining experimental concentration-dependent values of $IE_{vac}$, $IE_{EF}$, and resulting $e\Phi$s, the contributions of the latter to experimental spectral shifts can be quantified. We have been able to deconvolute previously observed changes in $VIE_{vac}$ into contribution of net surface molecular dipole changes (here, the electric double layer), and electronic-structure effects, caused by changes in the interactions between water molecules, which are typically weakened by additional solvent–solute interactions. In the present case of TBAI aqueous solutions, it is found that the surface effects dominate, with approximately 75% of the concentration-dependent $VIE_{vac}$ shifts being attributable to the interfacial dipole. The remaining 25% are thus associated with electronic-structure changes within the solution's interface, a result which is not obvious for a surfactant. The ability to quantify these important descriptors for solutions is a milestone in the characterization of solution–vacuum interfaces in general, both in terms of geometric and electronic structure. Generally, a detailed, microscopic interpretation of experimental data, like that presented here, will require dedicated theoretical modeling. Such computations are not yet available but will be triggered by the new ability to experimentally access quantitative and solute-concentration-dependent work functions from solutions. The present pioneering study also marks an advance towards connecting solution-phase electronic structure to electrochemistry, *e.g.*, by directly accessing a solution's equilibrium properties such as the electrochemical potentials of redox couples.

## Author Contributions

B.W. and S.T. conceived the experiments. M.P., B.C., I. Walter, S.M., F.T., D.S., U.H., and B.W., G. M. planned, prepared, carried out the experiments, and discussed the data. M.P., I. Walter, and S.T. analyzed the data. I. Wilkinson, B.W., and S.T. wrote the manuscript with feedback from all authors.

## Data Availability

The data of relevance to this study have been deposited at the following DOI: 10.5281/zenodo.7784822

## Conflicts of interest

There are no conflicts to declare.

## Acknowledgements

We acknowledge DESY (Hamburg, Germany), a member of the Helmholtz Association HGF, for the provision of experimental facilities. Parts of this research were carried out at beamline P04 of the synchrotron PETRA III, and we would like to thank Moritz Hoesch and his team for assistance in using P04. Beamtime was allocated for proposal II-20210015 (LTP). M.P., B.C., S.M., D.S., U.H., and B.W. acknowledge funding from the European Research Council (ERC) under the European Union's Horizon 2020 research and innovation program under Grant Agreement No. GAP 883759–AQUACHIRAL. U.H., S.M., and B.W. acknowledge support by the Deutsche Forschungsgemeinschaft (Wi 1327/5-1). F.T. and B.W. acknowledge support by the MaxWater initiative of the Max-Planck-Gesellschaft. B.C. acknowledges funding by the EPFL-MPG Doctoral School. S.T. acknowledges support from the JSPS KAKENHI Grant No. JP20K15229.

## Notes and References

N1  In the case of an anion, the respecitve energy would be the detachment energy, DE. We will refer to ionization energies throughout the text unless explicitly referring to anions.

N2  In some cases, such as for the non-bonding $1b_1$ HOMO ionization feature of gaseous water, the vertical IE happens to be equal to the adiabatic IE (AIE), *i.e.*, the minimum amount of energy required to remove an electron from a neutral molecule and populate the fully relaxed ionic ground state.

N3  In our previous aqueous TBAI solution study,[15] we have argued that the observed concentration-dependent $IE_{vac,1b1}$ changes primarily arise from $e\Phi$ changes. We will return to this point later, when discussing the origin of the concentration-dependent $e\Phi$ changes reported here.

N4  In the present case of a surfactant, the bulk and interfacial properties, such as the solute concentration, are necessarily different. Our 'bulk electronic structure' referrals are to the general case of solutions with similar bulk and interfacial solute-to-solvent concentration ratios, *i.e.*, where properties in the (probed) interface and deep into the bulk are equal. This term also includes the case of neat water without a solute, where electronic structure properties, such as VIEs, have been shown to be essentially invariant between the interface and the bulk.

N5  We note that PE-spectroscopic measurements can also be performed–from stationary droplets[26] or microdroplets,[48] which do not generate a streaming potential. However, these sample targets are nonetheless unsuitable for Fermi-referencing methods. For example, static droplets quickly accumulate surface contaminations[26] and free-floating microdroplets are not electrically connected to the apparatus ground potential. Thus, we will not further consider these methods here.

# Supporting Information

## How to measure work functions from aqueous solutions

Michele Pugini[1], Bruno Credidio[1], Irina Walter[1], Sebastian Malerz[1], Florian Trinter[1,2], Dominik Stemer[1], Uwe Hergenhahn,[1] Gerard Meijer[1], Iain Wilkinson[3], Bernd Winter[1]*, and Stephan Thürmer[4]*

[1] *Fritz-Haber-Institut der Max-Planck-Gesellschaft, Faradayweg 4-6, 14195 Berlin, Germany*
[2] *Institut für Kernphysik, Goethe-Universität, Max-von-Laue-Straße 1, 60438 Frankfurt am Main, Germany*
[3] *Institute for Electronic Structure Dynamics, Helmholtz-Zentrum Berlin für Materialien und Energie, Hahn-Meitner-Platz 1, 14109 Berlin, Germany*
[4] *Department of Chemistry, Graduate School of Science, Kyoto University, Kitashirakawa-Oiwakecho, Sakyo-Ku, 606-8502 Kyoto, Japan*

**Flow-Rate-Dependent Measurements of TBAI Aqueous Solutions**

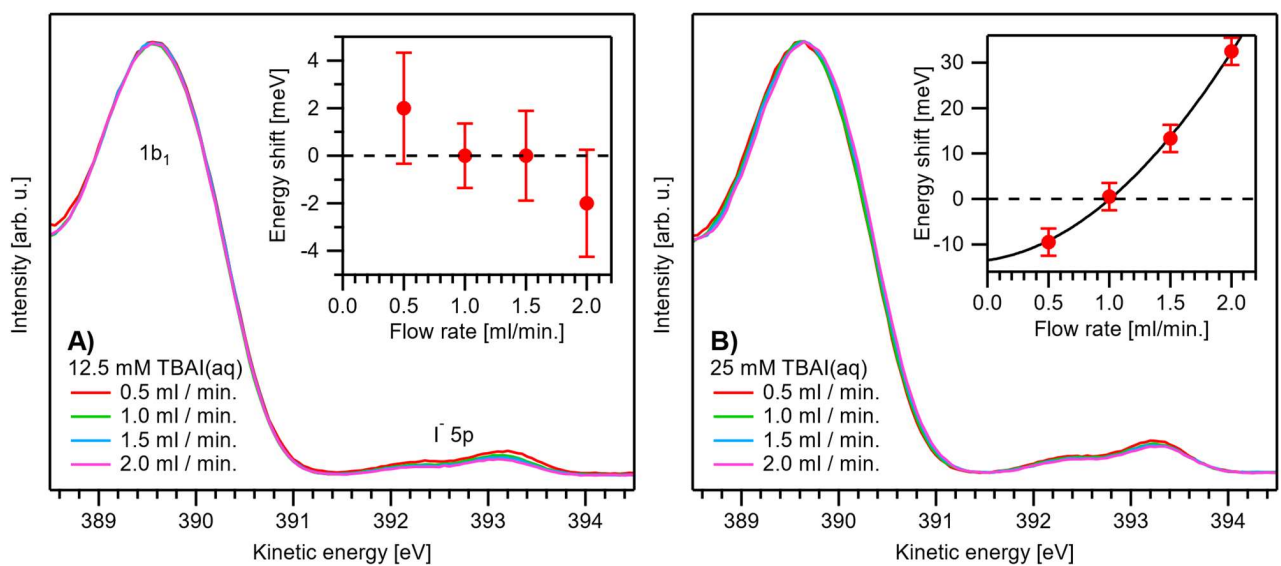

**Figure S1:** PE spectra of 12.5 mM **(A)** and 25 mM **(B)** TBAI$_{(aq)}$ showing the water 1b$_1$ and I$^-$ 5p PE features as a function of flow rate from 0.5-2.0 ml/min. Insets show the relative energy shift of the water 1b$_1$ feature compared to 1 ml/min flow. The error bars are fit errors. The polynomial fit in the inset of B) is only a guide to the eye.

PE spectra were recorded from aqueous solutions injected into the spectrometer at various flow rates and with representative concentrations of 12.5 and 25 mM TBAI, to assess their residual streaming potentials. Any energy shift in the spectra as a function of liquid flow rate can be extrapolated to zero flow, at which the streaming potential must vanish. Alternatively, should there be no shift, the absence of a streaming potential can be confirmed for any flow rate. Figure S1 shows the PE spectra for 12.5 mM (A) and 25 mM (B) solutions for flow rates from 0.5-2.0 ml/min. All spectra were recorded at the P04 soft X-ray beamline[49] of the PETRA III synchrotron facility (Deutsches Elektronen-Synchrotron, DESY, Hamburg) at a photon energy of 401 eV and using the same spectrometer setup (*EASI*) and PtIr microplate assembly as in the laboratory experiments. The insets show the energy shift of the water 1b$_1$ PE feature relative to the flow rate of 1 ml/min (dashed line), which was utilized for all measurements presented in the main text. For 12.5 mM TBAI, no significant shift and thus no notable contribution of the streaming potential is observed. For 25 mM TBAI, while showing a clear trend towards higher energy shifts at higher flow rates, the difference between an (extrapolated) zero flow and 1 ml/min is below 15 meV. This gives us confidence that the contribution of the streaming potential is an insignificant fraction of the energy shifts discussed in the main text.

## The Apparatus Fermi Level and Work Function

Measuring $E_F$ from *metallic* systems is straightforward. Since electronic states are occupied in conductors up to $E_F$, the associated energy is revealed as a sharp, high-energy cutoff in PE spectra. In the absence of any disturbing potentials, the measured kinetic-energy value associated with $E_F$ corresponds to $eKE_{EF} = h\nu - e\Phi_{det}$, with $e\Phi_{det}$ being the work function of the detection system. This somewhat counterintuitive result originates from the fact that photoelectrons experience the contact potential difference between the sample and apparatus, $\Delta e\Phi = e\Phi_{liq} - e\Phi_{det}$. The kinetic energy of the detected photoelectrons is thus modified by $\Delta e\Phi$, *i.e.*, $eKE_{EF} = h\nu - e\Phi_{liq} + \Delta e\Phi = h\nu - e\Phi_{det}$. For semiconductors and insulators – *i.e.*, condensed-phase systems exhibiting a band gap where the energy range around $E_F$ is devoid of available states for electrons – $E_F$ is not directly measurable. Similarly, liquid water does not exhibit electronic states in the vicinity of its Fermi level and can be modelled as a semiconductor with a direct band gap greater than 8 eV, where the exact value is still under debate.[50-52]

Note that the measured Fermi-level-feature kinetic energy, $eKE_{meas}$, is somewhat arbitrary, as it may be subject to experimental offsets associated with the detection system, such as a poorly calibrated energy scale. It is not an issue for Fermi referencing if the measured energy scale is linear, since all spectra are measured with the same device and are thus subject to the same systematic measurement offsets, *i.e.*, only relative kinetic energies are involved. This is the case for our hemispherical electron analyzer; care should be taken with non-linear systems such as time-of-flight analyzers, though. Previously, we corrected $eKE_{meas}$ with the offset factor for our system, determined as $0.224 \pm 0.008$ eV, which yielded $e\Phi_{det} = 4.293 \pm 0.009$ eV.[9] In the present experiments, we measured the 3p peak of Argon gas after the LJ experiments and compared it with the reference value of $15.759735 \pm 0.000001$ eV[53] to confirm the energy-scale correction. This yielded a consistent $e\Phi_{det} = 4.30 \pm 0.04$ eV; the somewhat larger error originates from observed fluctuations in the Ar 3p peak positions over multiple days.

## Pitfalls and Considerations for Accurate Fermi Referencing

While Fermi-referenced measurements seem straightforward, if streaming potentials and any other extrinsic potentials have been considered or eliminated, grounded measurements are prone to interferences from sub-optimal experimental conditions. Reliable measurements correspondingly necessitate the utmost care. We encountered several systematic experimental errors during our studies, which can negatively affect the measurement. Here, we give a brief overview of the associated potential pitfalls.

1) In the main text, it was noted that a consistent value was measured for the Fermi edge of our reference metal samples ($eKE_{EF} = 36.30$ eV), and that this value is stable, as it is an intrinsic property of our apparatus. We observed, however, that this value can be temporarily altered slightly – on the order of 50 meV or more – when measuring certain samples, such as solutions containing organic compounds. This may stem from persistent surface-adsorption layers on the inner walls of the apparatus, which desorb over the course of days or even weeks. We correspondingly recommend the constant monitoring of the Fermi-edge position when attempting to measure Fermi-referenced spectra.

2) Another important aspect in LJ-PES is the continuous evaporation from the liquid surface, which forms some rather ill-defined adsorbate layer on all inner walls of the apparatus. This alters the surface potentials of the interaction chamber, which causes spectral shifts over time and with varying chamber pressure; for example, a shift of several hundred meV has been observed over a 2–4-hours period after starting experiments.[9]

3) In the case of the glass-nozzle assembly, grounding was achieved via metallic inset tubes placed in between the HPLC pump and PEEK liquid delivery line prior to injection into the vacuum chamber. As flow characteristics of the solution in this metallic inset are altered, crystallization of solutes may occur, which deteriorates the grounding. We observed shifts on the order of ~100 meV of all PE spectra in such a case. Cleaning and reassembling the liquid delivery system recovered the original energetic positions of PE spectra.

4) Figure S3 compares PE spectra from the PtIr microplate, a new, passivated glass nozzle, and a glass nozzle after running solutions with a particularly high (or low) pH value. Large shifts (up to ~300 meV) of all PE spectra were observed in the latter case. The unshifted PE spectra could not be recovered, even after several days. It is very likely that the surface properties of the inner glass-nozzle walls were chemically altered, which could lead to a significant streaming potential, or alternative additional unwanted potentials. Furthermore, the metallic inset tube used to ground the solutions may have been corroded. Thus, we found that the glass-nozzle assembly is more prone to detrimental effects, and measurements from a PtIr microplate (or other nozzles which provide both proper grounding and chemical resistance) are preferred.

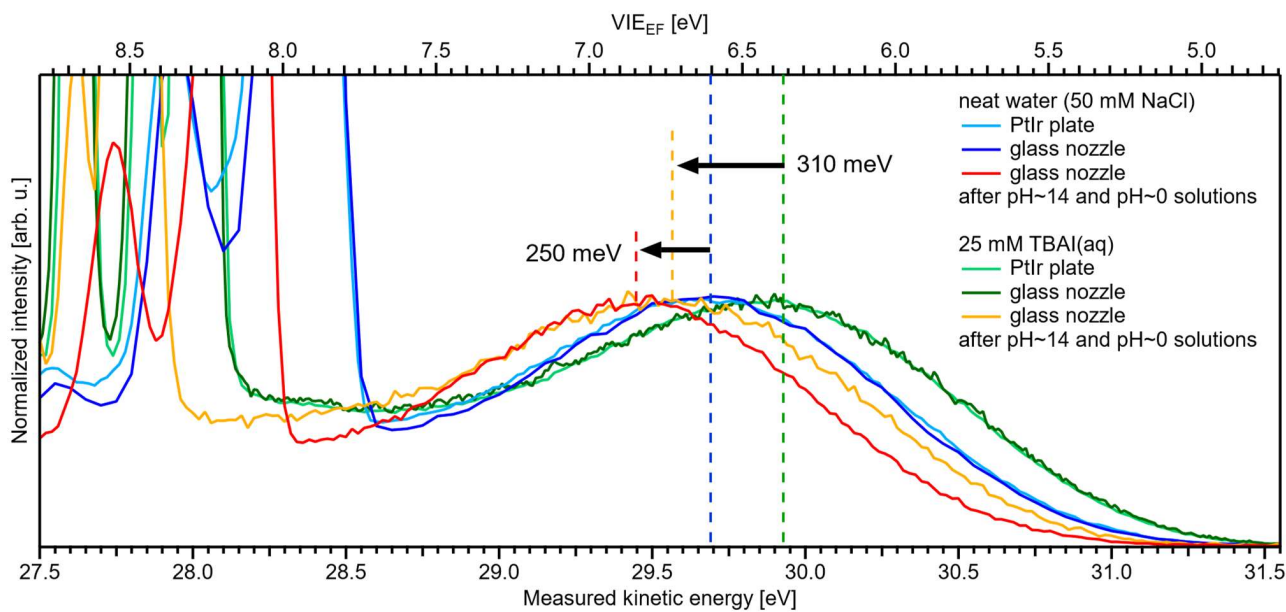

**Figure S2:** Exemplary valence PE spectra of reference water (with 50 mM NaCl added) and 25 mM TBAI$_{(aq)}$, measured from a grounded liquid jet with hv = 40.814 eV. If properly prepared, PE spectra measured from a PtIr microplate are equivalent to the ones from a glass capillary for both reference water (light and dark blue curves) and TBAI$_{(aq)}$ (light and dark green curves). However, the PE spectra shifted considerably after running solutions with a pH value well removed from 7; both a pH close to 0 and 14 was reached before the spectra shown here were measured. Specifically, shifts of ~250 meV and 310 meV were observed for reference water (red curve) and 25 mM TBAI$_{(aq)}$ (yellow curve), respectively. We noticed such shifts whenever solutions with particularly

high or low pH values were measured prior to Fermi-referenced measurements with glass-capillary nozzles, suggesting a pH-induced deterioration of the glass and an associated change in liquid streaming potential.

## Additional Figures

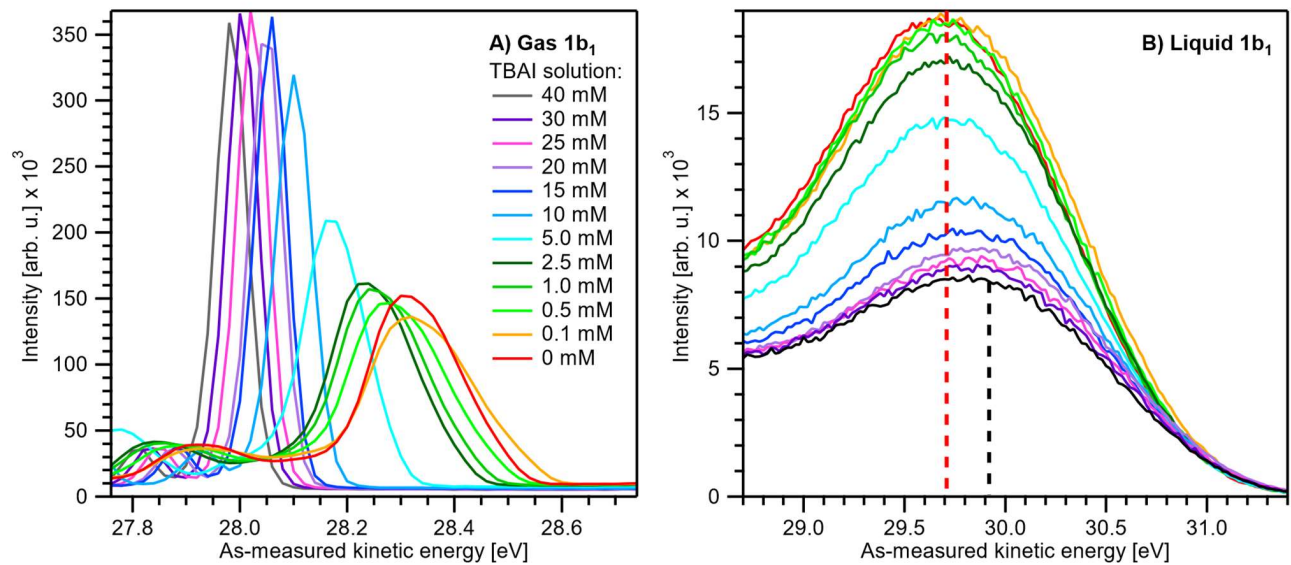

**Figure S3:** Close-up on the **A)** gas and **B)** liquid $1b_1$: The data from Fig. 4 in the main text are shown without intensity normalization and vertical offsets, *i.e.,* here, the PE intensities are shown as-measured.

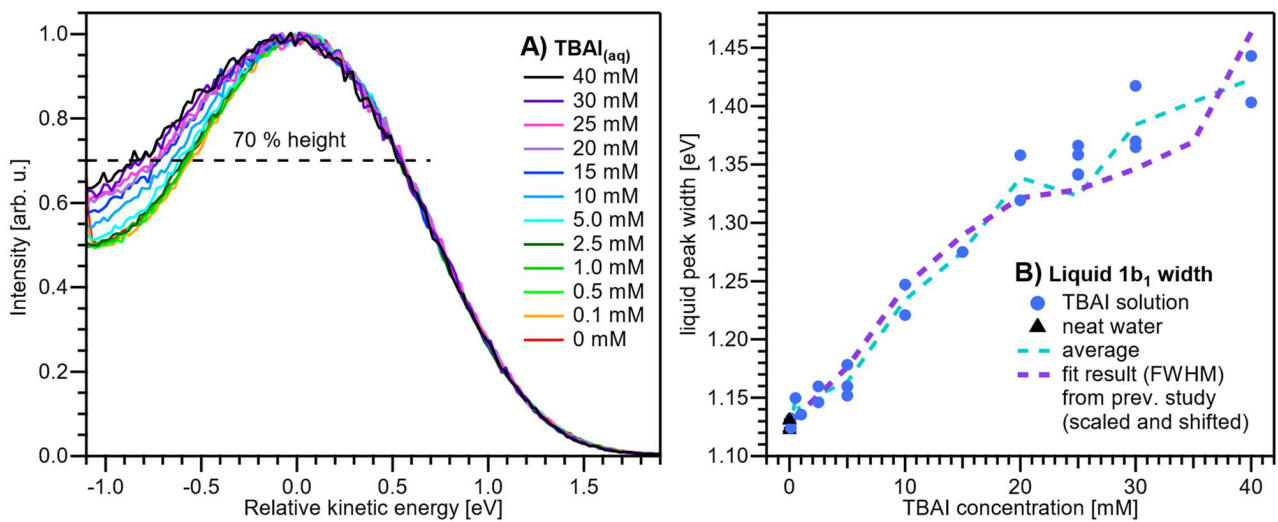

**Figure S4: A)** Liquid $1b_1$ band shape as a function of TBAI concentration; the spectra are the same as in Figs. 4 and S3. Spectra have been normalized to unit height at the $1b_1$ band and shifted to maximum overlap at the high-energy shoulder to emphasize the change in shape. **B)** Values of the peak width as extracted from A) as a function of TBAI concentration. The widths were taken from the $1b_1$ band at 70% signal height (dashed line in panel A) and do not originate from the fits discussed in the main text. Fit results (FWHM values) from our previous study[15] have been added for comparison; this data has been arbitrarily scaled by a factor of 3 and aligned to the reference water value.

In the main text, it was explained that the liquid water 1b$_1$ band was fitted by a single peak over a limited data range because of overlapping features from the gas phase, which made a full fit of the spectrum difficult. Thus, it was not possible to extract a peak width from the fit, due to the constrained nature of the procedure. Instead, in Fig. S4, we plot the total observed width of the PE band in the data directly at 70% band height, a value arbitrarily chosen to avoid interference from the signals of the nearby orbitals and the gas phase. The result in Fig. S4B serves as a confirmation of the overall trend of a broadening liquid-water 1b$_1$ band, as observed in our previous study,[15] which is also replicated as a purple dashed line. This confirms that the shape of the liquid bands is the same, both in the grounded and biased measurements. Any presence of $V_{tot}$ does not lead to additional broadening, since, similarly to the biased case, this potential leads to a rigid shift of all liquid features. This is furthermore consistent with the observed excellent match of biased and unbiased PE spectra; see, for example, Fig. 7A in the main text.

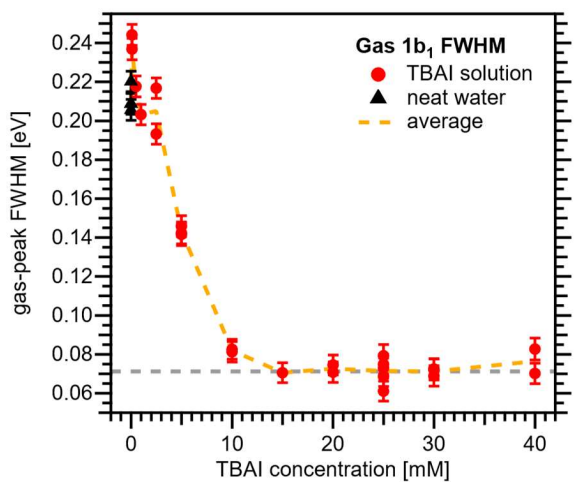

**Figure S5:** Peak width (FWHM) of the gas-phase 1b$_1$ peaks as a function of TBAI concentration from peak fits to the PE spectra. Error bars represent the quadratic addition of fit errors (one sigma) and a general uncertainty in the measurement of 5 meV. Red dots indicate results from TBAI$_{(aq)}$ (5 mM NaCl have been added for concentrations at and below 1 mM TBAI to assure sufficient conductivity) and black triangles indicate reference water results (no TBAI but 50 mM NaCl). The minimal FWHM (indicating $|V_{tot}| \approx 0$ V) is reached from about ~10 mM up to the highest concentration. The constant value of ~70 meV indicates that any changes are below our ability to measure peak widths in our experiment.